\documentclass[letterpaper,12pt]{article}
\usepackage{tabularx} % extra features for tabular environment
\usepackage{amsmath}  % improve math presentation
\usepackage{graphicx} % takes care of graphic including machinery
\usepackage[margin=1in,letterpaper]{geometry} % decreases margins
\usepackage{cite} % takes care of citations
\usepackage[final]{hyperref} % adds hyper links inside the generated pdf file
\usepackage{authblk}
\hypersetup{
	colorlinks=true,       % false: boxed links; true: colored links
	linkcolor=blue,        % color of internal links
	citecolor=blue,        % color of links to bibliography
	filecolor=magenta,     % color of file links
	urlcolor=blue         
}
\usepackage{blindtext}
\begin{document}

\title{ATMOSPHERIC NOISE MEASUREMENTS IN THE GARDEN: DETECTING UNIVERSALITIES IN INTER-SPHERIC WAITING TIME STATISTICS}
\author[1,*]{Istv\'an Gere}

\affil[1]{Department of Physics, Babeş-Bolyai University, Cluj-Napoca, Romania}
\affil[*]{email: istvan.gere@ubbcluj.ro}
\date{\today}
\maketitle

\begin{abstract}
Lightning flashes result in an instantaneous emission of electromagnetic (EM) waves that encompass a broad spectrum of frequencies in the domain of radio waves. The signature of these impulses in the region of the very low frequency (VLF) of radio waves and bellow are called spherics. These impulsive signals traverse great distances in the waveguide between the Earth's ionosphere and its surface. Due to the abundance of lightnings the lower end of the EM spectrum is dominated by the waveforms of these emission. It is called atmospheric radio noise because of its origin. This article outlines a straightforward approach for capturing spherics activity within the VLF radio wave range. Employing data processing techniques, we pinpoint the timestamps of spherics within time series data. The distribution of inter-spheric times is analyzed across various detection threshold levels. Utilizing recordings spanning two distinct years and seasons, we replicate the established form of the inter-spheric time distribution described in the literature. Notably, we demonstrate that by rescaling the time intervals between spherics with the mean time for a specific recording and given detection threshold, the distributions collapse to a master curve. This universal pattern is accurately characterized by a single-parameter mean-scaled Gamma distribution. Additionally, we note the similarities in the distribution of inter-spheric times with patterns found in earthquake recurrence times.

\end{abstract}

\newpage

\section{Introduction}

Lightning flashes produce a sudden emission of electromagnetic (\textbf{EM}) waves across a broad domain of the EM spectrum within the radio frequency range \cite{lightning1}. Studies on the radio frequency (RF) spectrum of these sudden emissions indicate that the peak of the spectral energy is typically around 10 kHz and decreases with the reciprocal of the frequency \cite{lightning_0}.The peek of the emitted energy overlaps with the range of very low frequency (VLF) ($30$ kHz - $3$ kHz) radio waves.In the lower ranges of the RF spectrum the cavity between the Earth’s surface and the ionosphere acts as a low-attenuation (around $2$-$3$ dB/$1000$ km) \cite{lightning2} waveguide enabling these instantaneous wide bandwidth EM emissions to travel great distances ($1000$-$2000$  km) \cite{lightning3}. The attenuation further decreases during the night as a result of the diurnal cycle, as radiation from the Sun does not reach certain regions of the Earth. This restructuring of electron density in the upper layers of the ionosphere creates even more favourable conditions for propagation \cite{lightning4}. In the literature, these electromagnetic emissions induced by lightning are referred to as radio atmospheric signals or, more concisely, as \textit{spherics} \cite{lightning5}. The global lightning activity is measured in lightning flashes per second. Recent estimation methods based on satellite data estimate the global yearly average to be $44\pm5$ flasher per second, with hemispheric, local and seasonal variations \cite{lightning6}. The abundant number of spherics generated by global lightning activity occupy the RF spectrum, contributing to atmospheric electromagnetic noise. Specially designed radios make it possible to record this noise. In this article, we investigate the VLF spectrum of spheric emissions for two main reasons: firstly, because the VLF spectrum of radio waves aligns with the audible frequency range of sound, making these signals easily audible with simple amplification; and secondly, because there are readily available data acquisition tools designed specifically for this frequency range, such as computer sound cards and high-resolution dictaphones.The fact that the peak of the lightning spectral energy is also found in this range of VLF domain and that the attenuation is low at night further enhances the reception of the spherics \cite{lightning_0,lightning2}. In this paper we  present a simple setup of a special radio that allows us to intercept and record spherics originating from remote thunderstorms.We present an example spectrogram of such recordings. The distribution of inter-spheric times for various detection thresholds, as calculated using the procedure outlined in \cite{Poissons}, replicates their findings.

Contrary to the complex description provided in \cite{Poissons}, we offer a simpler account. Drawing inspiration from our earlier studies on earthquakes \cite{earthquake}, a noteworthy observation emerges: by rescaling the time between spherics, denoted as $t$, with the mean time $\langle t \rangle$ for a particular dataset of inter-spheric times (across different years and under various $L$ detection thresholds) to obtain $\tau = \frac{t}{ \langle t \rangle}$, the experimental probability densities, represented by $\rho(\tau)$, collapse to a common master curve.  This universal curve can be described as a single parameter Gamma-distribution.

\section{Experiment}
\subsection{The VLF radio device}
\begin{figure}[h!tb]
\centering
\includegraphics[width=0.6\textwidth]{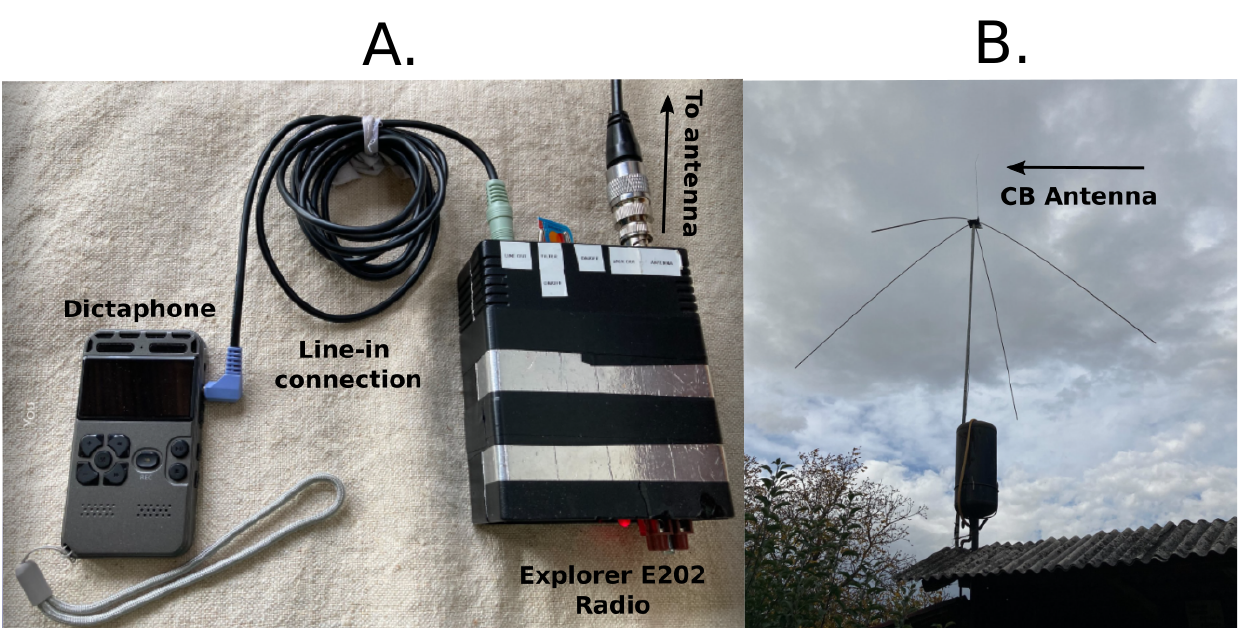}
\caption{\textbf{A.)} Experimental setup featuring the Explorer E202 connected to both the dictaphone and the antenna cable. \textbf{B.)} The CB Antenna mounted on a metal rod with four ground plane radials (suboptimal for VLF signals, but performed well in our experiments due to its height and location).}
\label{pic1}
\end{figure}

To perform our experiments we used the Explorer E202 broadband radio specifically designed for VLF spherics reception. Open-source schematics of the Explorer E202 \cite{Explorer232} with off the shelf electric components were used to build this device, which is presented in Figure \ref{pic1}. The core of this device is a TL081C  operational amplifier \cite{TLC} that is connected to the antenna input through an impedance matching circuit. The device includes two outputs, a line-level type for recording and a headphone level type for real time field listening \cite{Explorer232}. The device operates on a single 9V battery, making it portable while it ensures a stable current source and low internal electric noise.

The experimental setup involved a citizens band (CB) radio antenna vertically positioned at approximately 4 meters in height (refer to Figure \ref{pic1} B.). This antenna was connected to the radio through a coaxial cable (see Figure \ref{pic1} A.). Additionally, the device was linked to a dictaphone with high-resolution recording capabilities using a simple two-way audio jack (see Figure \ref{pic1} A.). The dictaphone is preferred over directly using a laptop, as it enables longer recording sessions and is free from the internal electronic noise associated with laptops. It is important to note that, according to the specifications of the Explorer E202, the CB antenna used in our setup is sub-optimal. A longer vertical antenna could have increased the signal-to-noise ratio, although this did not pose significant issues during the recordings, due to good reception conditions.

\subsection{Data acquisition}

The recordings were performed using the dictaphone, in our case serving as an analog-digital converter unit (ADC-converter), configured in a line-in mode that is compatible with the line-out mode of the Explorer E202. A consumer line-level is a standard in analog audio transmission with a peak-to-peak voltage of $V_{pp} = 0.894\textrm{ V}$  \cite{line_level}. The ADC-unit digitalizes these signals with a sample rate $\nu_s = 48$ kHz and a bit depth of $16$ bit per sample. According to the Nyquist-Shannon sampling theorem \cite{shannon}, this allows frequencies to be representable up to $\frac{\nu_s}{2}$  frequencies, which overlaps with the studied VLF frequency range.

\begin{figure}[h!tb]
\centering
\includegraphics[width=0.7\textwidth]{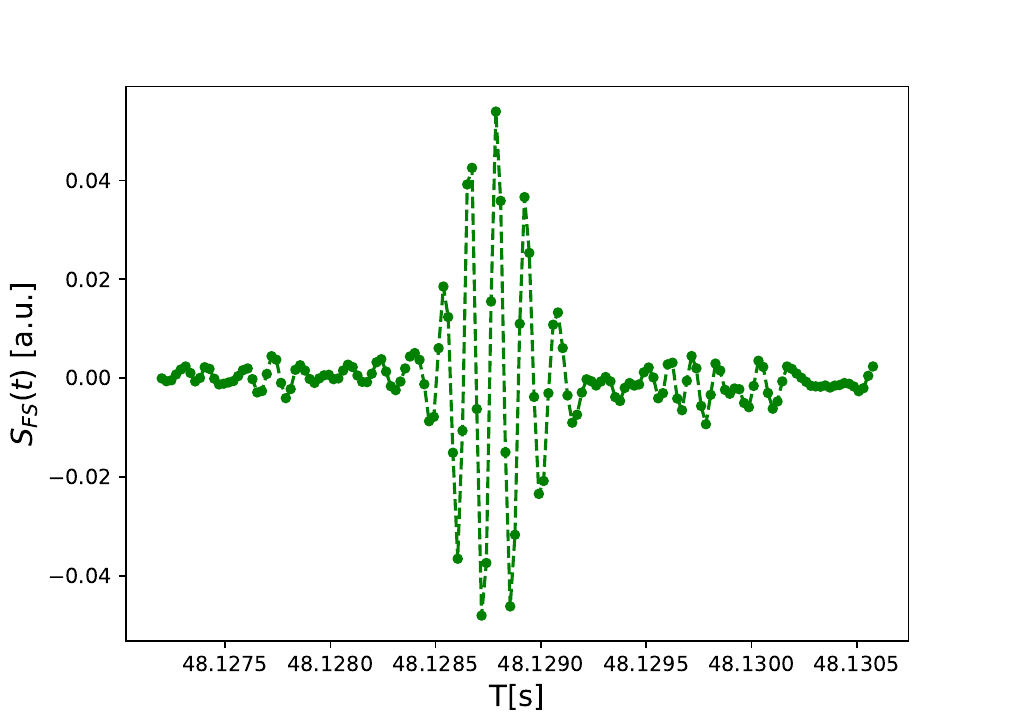}
\caption{The waveform of a single spheric in time.}
\label{pic2}
\end{figure}

\begin{figure}[h!tb]
\centering
\includegraphics[width=0.9\textwidth]{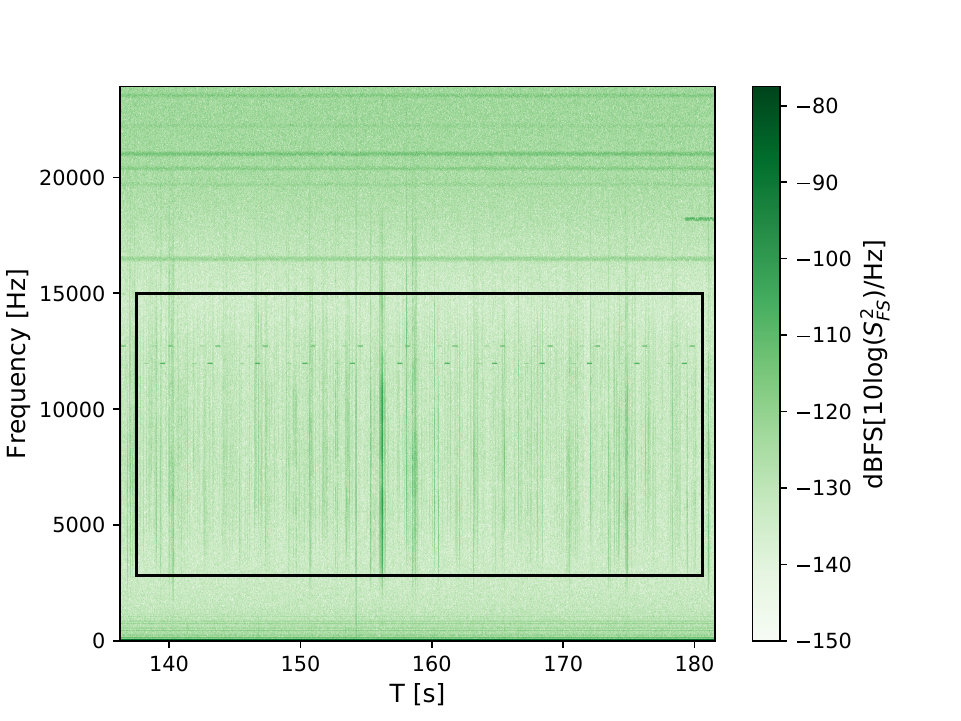}
\caption{The spectrogram of a $\approx$ 40 seconds excerpt from a recording performed in 2022. The vertical lines represent the signature of spherics. The power spectral density values, denoted as $\textrm{dBFS}[10\textrm{log}(S_{FS}^2)/\textrm{Hz}]$ are calculated based on the signal strength relative to the full scale $S_{FS}$ using a Fast Fourier Transform (FFT), with a time windowing of 2042 samples. At the lower part of the spectrogram the harmonics of the 50 Hz power line hum is visible. The vertical lines between $11$ kHz and $13$ kHz, at regular intervals are the signals of the Alpha (or RSDN-20) navigation \cite{alfa}. The black rectangle represents the frequency range of $4$ to $15$ kHz, the frequency domain in which we investigate spheric phenomena..}
\label{pic3}
\end{figure}

In Figure \ref{pic2}, we provide an illustration of the waveform of a single spheric. The signal strength over time is computed as follows: let $\textrm{s}(t)$ denote the value detected by the Analog-to-Digital Converter (ADC) using a 16-bit notation, and $FS$ (full scale) represent the maximum representable value in the 16-bit notation, given by $FS = 2^{16}-1 = 65535$. The absolute signal strength relative to $FS$ is then expressed as $S_{FS}(t) = \frac{\textrm{s}(t)}{FS/2}$. The $S_{FS}(t)$ quantity is proportional to the electric field strengths detected by the antenna. The relative signal strength $S_{FS}$ is sufficient for our study as we are looking for the time signatures of the spherics, to calculate the inter-spheric time. In Figure \ref{pic3} we presents the spectrogram of a short signal sample.

\begin{figure}[h!tb]
\centering
\includegraphics[width=0.9\textwidth]{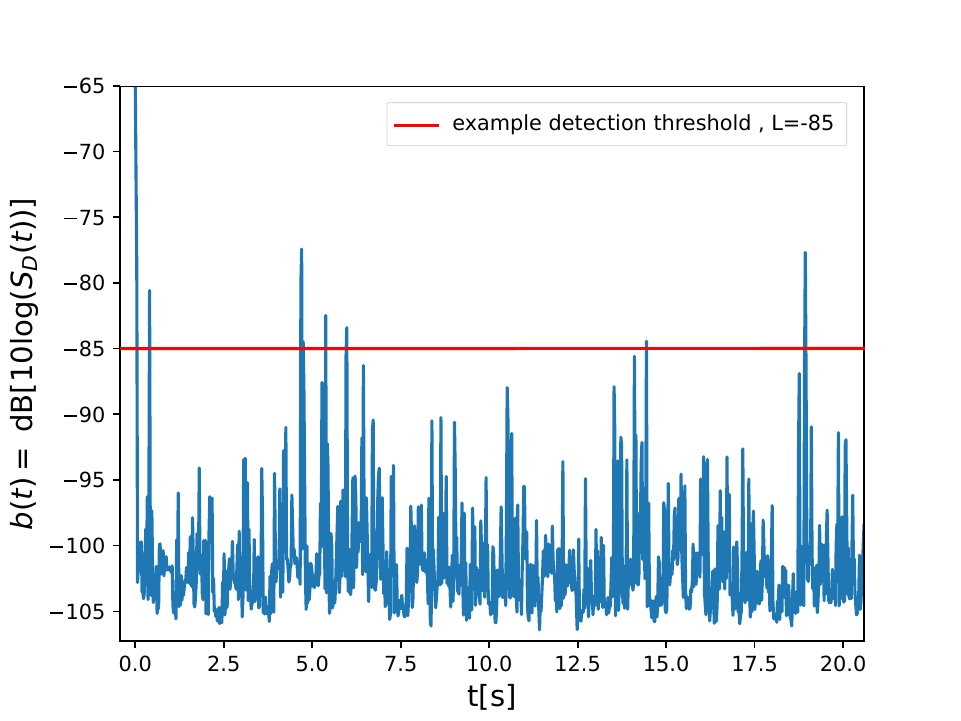}
\caption{On this plot we present the measure $b(t) = \textrm{dB}[10\textrm{log}(S_D(t)]$ employed for spheric detection. In this presentation the prominent spikes correspond to lightning spherics. The red line at $L = -85$ dB presents an example threshold to detect spherics, whereas peaks over this line are considered spherics.}
\label{pic4}
\end{figure}

As indicated in the highlighted area of Figure \ref{pic3}, the spectral signature of spheric phenomena in our recordings is observed between $1$ kHz and $15$ kHz. The spectrograms are generated using Fast Fourier Transform (FFT) applied to a sample window of $2024$ samples centered around a time-step $t$ in the digitalized data, with a $50\%$ overlap. The next step involves extracting a measure of spectral strength within the frequency domain $D = [4 \textrm{ kHz}, 15\textrm{ kHz} ]$. We accomplish this by summing the power spectral density of frequencies within the domain $D$ at each time-step $t$, expressed as $S_D(t) = \sum_{D}S_{FS}^2(\nu)_t$. This measure, $S_D(t)$, is proportional to the power spectral density of the frequency domain $D$ at the midpoint $t$ of the FFT sample window. Using this method, we effectively eliminate electromagnetic noise from the lower end of the spectrum, originating from the power grid's $50$ Hz signal and its harmonics, commonly referred to as mains hum \cite{hum}. Additionally, we remove noise from the upper end of the spectrum, which does not contain relevant information for this study. From the spectral strength $S_D(t)$ over the frequency domain $D$ the following measure is constructed $b(t) = \textrm{dB}[10\textrm{log}(S_D(t)]$ We present it on Figure \ref{pic4}.

Using the new measure $b(t)$ shown in Figure \ref{pic4} lightning spherics can be readily identified along with their precise timing. When the measure $b(t)$ surpasses a predefined threshold level $L$, the exact time $t$ when the measure reaches a local maximum is located using a peak finder algorithm\cite{peakfinder}. Applying this detection method, we generated datasets containing the precise times of each detected lightning spheric by our device. The recording location is situated the in the village of Sâncraiu, Romania (46.8304° N, 22.9893° E) chosen for its rural setting with lower artificial EM noise levels, making it an optimal environment for recordings. Our experimental measurements encompassed two recording sessions:
\begin{enumerate}
 \item In 2022-07-04, starting at 21:34:46, a 50 min recording.
 \item In 2023-04-22, starting at 19:40:53, a 80 min recording.
\end{enumerate}
%Although, such a simple device does not allow us to infer the absolute strength of lightning, we can still study the properties of this atmospheric noise that arises from the lightning activity.

\section{Results}
In this section we reflect on the probability distribution of inter-spheric times, as observed in the experimental recording sessions processed using the methods detailed earlier. A representative example of spheric detection threshold is depicted in Figure \ref{pic4} where the red line delineates the threshold beyond which a peak is deemed a spheric.By reducing the threshold, we incorporate a greater number of peaks, encompassing weaker spheric signatures. By applying different threshold levels to the measure $b(t)$, we created multiple datasets containing the inter-spheric times for the two experimental sessions. In Table \ref{table1} we present the detection threshold $L$ with the corresponding the number of detected spherics $N$, and the mean time between spherics $\langle t \rangle$. For each threshold $L$ and year, there corresponds an inter-spheric time database, as presented in Table \ref{table1}.

\begin{table}[h!t]%
\caption{The table displays the observed number of spherics $N$, the mean time between two spherics $\langle t \rangle$, and the corresponding threshold level of detection $L$.}
\centering
\begin{tabular}{|c|c|c|c|}
\hline
\begin{tabular}[c]{@{}c@{}}Measurement\\ session\end{tabular} & \begin{tabular}[c]{@{}c@{}}Threshold \\ $L$ {[}dB{]}\end{tabular} & \begin{tabular}[c]{@{}c@{}}Number of Spherics\\ detected $N$\end{tabular} & \begin{tabular}[c]{@{}c@{}}Mean time between\\ spherics\\ $\langle t \rangle${[}s{]}\end{tabular} \\ \hline
{}{}{I. 2022}                                      & -83                                                               & 810                                                                       & 3.7                                                                                               \\ \cline{2-4}
                                                              & -85                                                               & 1108                                                                      & 2.71                                                                                              \\ \cline{2-4}
                                                              & -89                                                               & 2149                                                                      & 1.39                                                                                              \\ \cline{2-4}
                                                              & -92                                                               & 3511                                                                      & 0.85                                                                                              \\ \cline{2-4}
                                                              & -96                                                               & 6733                                                                      & 0.44                                                                                              \\ \hline
{}{}{II. 2023}                                      & -83                                                               & 162                                                                       & 18.3                                                                                              \\ \cline{2-4}
                                                              & -85                                                               & 233                                                                       & 12.72                                                                                             \\ \cline{2-4}
                                                              & -90                                                               & 661                                                                       & 4.5                                                                                               \\ \cline{2-4}
                                                              & -92                                                               & 999                                                                       & 2.99                                                                                              \\ \cline{2-4}
                                                              & -95                                                               & 2644                                                                      & 1.13                                                                                              \\ \hline
\end{tabular}

\label{table1}
\end{table}

%%%%%%HERE%%%%%%
%%%%%%Here%%%%%%

\begin{figure}[h!tb]
\centering
\includegraphics[width=1.\textwidth]{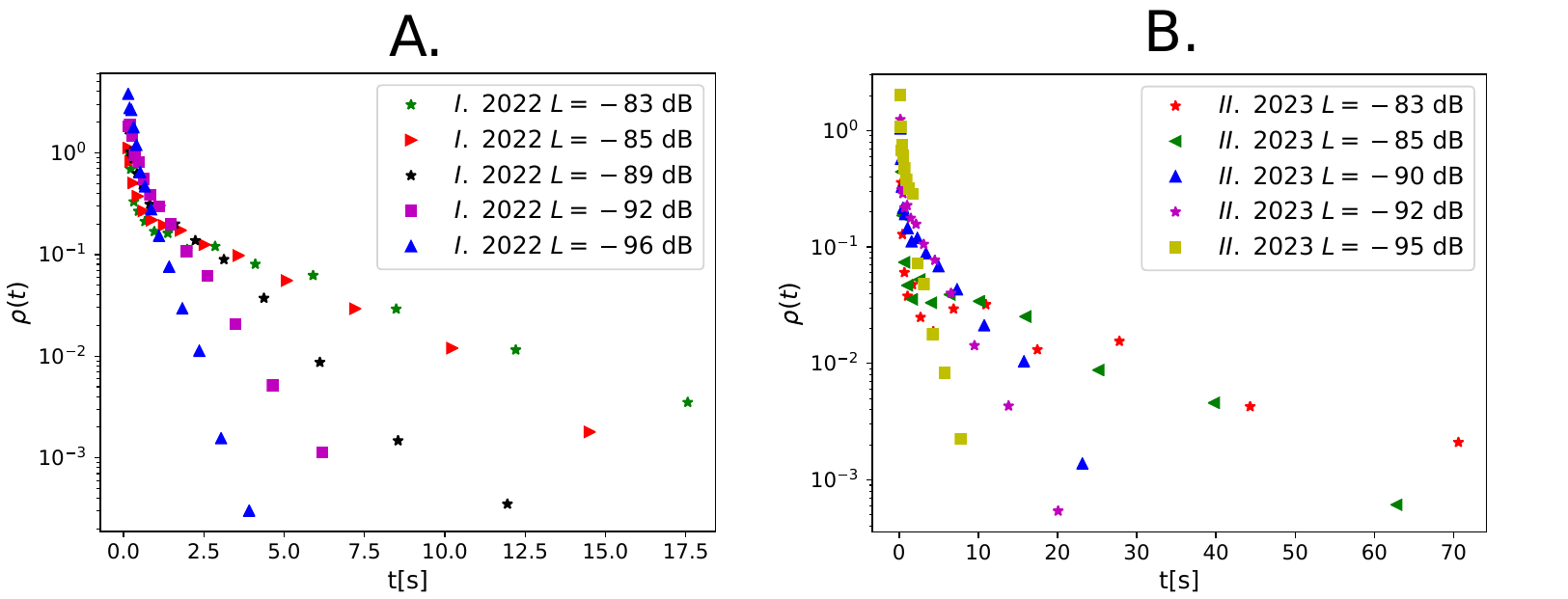}
\caption{The experimental probability density of inter-spheric times for the 2022 (\textbf{A.}) and 2023 (\textbf{B.}) measurements, considering various thresholds as described in Table \ref{table1}. For $t > 1.5$ s, the distributions asymptotically converge towards an exponential pattern over time.}
\label{pic5}
\end{figure}

In Figure \ref{pic5} we present the probability density of the inter-spheric times observed in our experiment with different $L$ thresholds of spheric detection from the datasets with the properties found in Table \ref{table1}. Based on the observed shapes of these experimental probability densities, we infer that, for inter-spheric times exceeding 1.5 seconds, the distribution tends to converge towards an exponential distribution (indicated by a straight line on a semi-log plot). This behavior aligns with similar patterns depicted in plots found in the existing literature \cite{Poissons}. The exponential characteristic of the tail is attributed to the detection of lightning occurrences from multiple independent storms around the world at our recording location.

\begin{figure}[h!tb]
\centering
\includegraphics[width=1.0\textwidth]{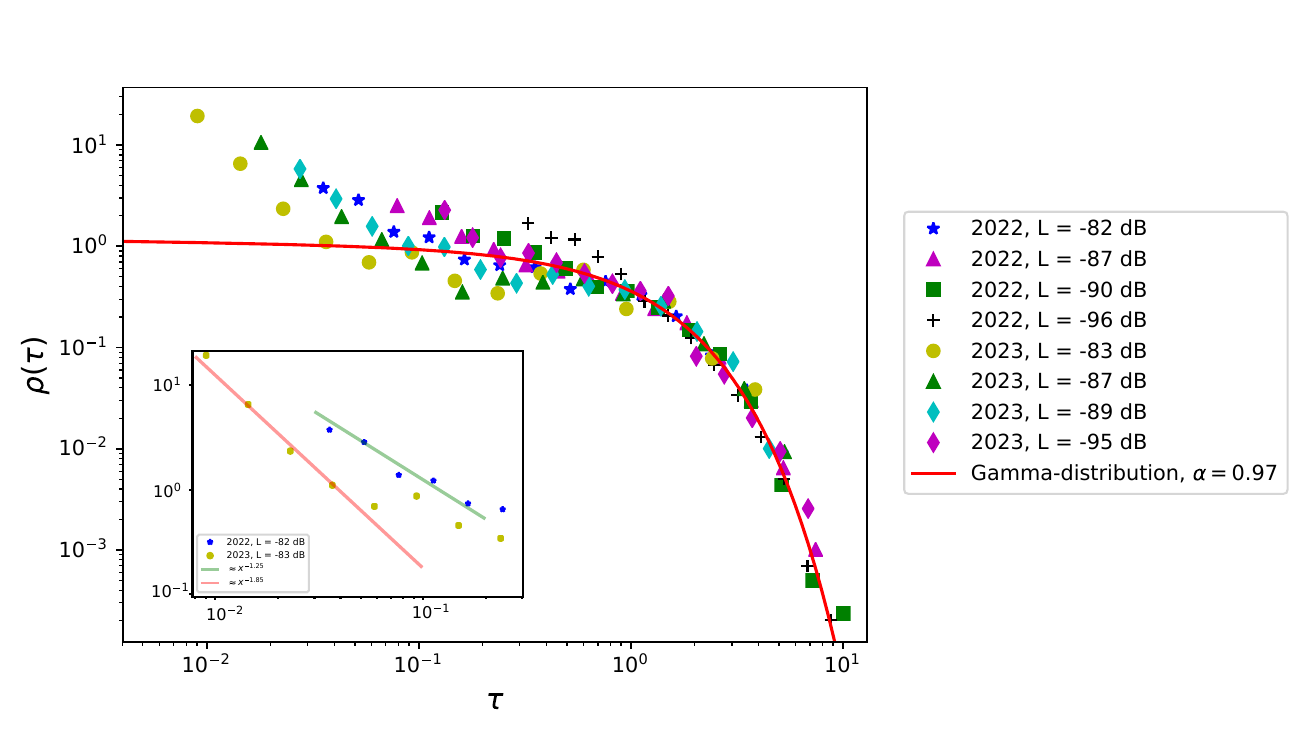}
\caption{The plot illustrates the probability densities $\rho(\tau)$ of mean-rescaled inter-spheric times $\tau$, at various threshold levels and across different years. The observed master curve can be described with the mean-rescaled Gamma distribution in the form of equation (\ref{eq1}) with the parameter $\alpha = 0.97$. The inset figure provides a detailed look at the lower segment of the plot corresponding to 2022 and 2023 datasets, where the distribution tends towards a power law, exhibiting exponents of $-1.85$ for 2022 and $-1.25$ for 2023. }
\label{pic6}
\end{figure}

After obtaining the time distribution of inter-spheric intervals, our next objective was to characterize our findings using a statistical physics approach, a methodology already applied in geology to describe the recurrence time of earthquakes \cite{earthquake}. Unlike the prevailing literature, which relies on a highly parametrized model, our approach seeks simplicity and universality in capturing the underlying dynamics of the observed inter-spheric time \cite{Poissons}. For the time datasets created through thresholding (with properties detailed in Table \ref{table1}), we constructed the experimental probability density of the mean-scaled inter-spheric times using the mean time of the dataset, denoted as $\tau = \frac{t}{\langle t \rangle}$. The rescaled probability densities, denoted as $\rho(\tau)$, are depicted in Figure \ref{pic6}. We observe a novel universality, where, for different measurement sessions and threshold levels, the rescaled distributions collapse onto the same master curve. This curve can be described by the mean-rescaled Gamma distribution, as represented in equation (\ref{eq1}). The presence of the master curve and the single-parameter Gamma distribution, describing the recurrence time in earthquakes and also observed in the distribution of time between acoustic emissions during metal deformation (compression of micro-capillaries), has been documented in the literature \cite{earthquake}. This suggests universality in the processes producing these diverse natural phenomena.

\begin{equation}
 \rho(\tau) = \frac{\alpha^{\alpha}}{\Gamma(\alpha)} \tau^{\alpha - 1} e^{-\alpha \tau}
\label{eq1}
\end{equation}

For $\tau < 0.1$ we also observe a power-law like decay to the Gamma distribution. A comparable power-law like decay in recurrence time toward the Gamma distribution has also been observed in the previously mentioned study of earthquakes \cite{earthquake}. In such cases, the empirical Omori law \cite{omori} is often considered to be the source of these deviating regions in inter-event time distribution. In seismology it describes the number of aftershocks (clustering) after a main event as a power law of the time with a negative exponent. The clustered set of events in the experimental inter-event times manifests as a short region with low $\tau$ values, which can be approximated with a power law \cite{omori_in_series}. We present such cases selected from the experimental probability densities for spherics in the Figure \ref{pic6} inset plot, where the regions dominated by a power-law are highlighted. The inherent clustering observed in the occurrence of spherics originates from the nature of lightning events. Typically, lightning is not a singular occurrence; instead, it comprises an initial main lightning followed by multiple flashes. In this sequence, the same ion channel from the initial event conducts additional charge, leading to subsequent flashes \cite{Poissons}.

\section{Conclusions}
In this study, we introduced a simple methodology for capturing the radio waves produced by lightning events within the very low-frequency range. The atmospheric electromagnetic noise that characterizes this frequency range can be studied for its statistical characteristics. We studied the temporal behaviour of the spherics, electromagnetic signatures of remote lightnings. These spherics signals are present as a sharp and sudden increase in signal strength over the noise limit of the recordings. Processing data from two recording sessions conducted in different years and seasons, we generated datasets of inter-spheric times with varying detection thresholds. We observed that rescaling individual inter-spheric times with the mean inter-spheric time of the dataset resulted in collapsed probability distributions onto a common master curve. This curve can be accurately described by a mean rescaled Gamma distribution, as indicated by equation (\ref{eq1}) with a single parameter $\alpha \approx 0.97$. To our knowledge, this universality has not been documented in the literature. Our one-parameter fit provides a more straightforward description of the observed inter-spheric time distribution compared to the results reported in existing literature. We also note a resemblance in the distribution patterns of inter-spheric times to earthquake recurrence times, and the intervals between acoustic emissions during the deformation of metals \cite{earthquake}. While all of these phenomena can be modeled with a Gamma function, we also observed a distinct section with low values in the density distributions, exhibiting high temporal correlations. In these regions, the distribution tends to converge towards a power-law. This implies a universality in the underlying processes governing these diverse natural phenomena, placing spherics within the broader family of such phenomena.

\subsection*{Acknowledgements}
This work supported by the Romanian UEFISCDI PN-III-P4-ID-PCE-2020-0647 research grant. I would also like to express my gratitude to O. Botos for his guidance on the implementation of the Explorer E202 radio. Special thanks to Z. I. Lazar for his invaluable advice throughout the writing of this article.

%\newpage

\end{document}